\documentstyle[amssymb,11pt]{article}

\global\arraycolsep=1pt 
\input{tcilatex} 
\input epsf 
\begin{document}

\hfill CERN-TH/99-383

\medskip

\medskip

\medskip

\medskip

\begin{center}
\vspace{10pt}{\Large {\bf IRREVERSIBILITY AND HIGHER-SPIN \\[0pt]
CONFORMAL\ FIELD\ THEORY }}

\medskip

\medskip

\medskip

{\sl Damiano Anselmi}

{\it CERN, Theory Division, CH-1211, Geneva 23, Switzerland}

\medskip

\medskip

\medskip

\medskip

\medskip

\vspace{12pt} {\bf Abstract}
\end{center}

\begin{quote}
\vspace{4pt} {\small I discuss the 
properties of the central charges $c$ and $a$ for
higher-derivative
and higher-spin theories (spin 2 included). Ordinary 
gravity does not admit a straightforward identification
of $c$ and $a$ in the trace anomaly, because it is not
 conformal. 
On the other hand, higher-derivative theories can be conformal, but
have negative $c$ and $a$. 
A third possibility is to consider
higher-spin conformal field theories. 
They are not unitary, but have a variety of interesting properties.
Bosonic conformal tensors have a
positive-definite action, equal to the square of a field strength, and a
higher-derivative gauge invariance. There
exists a conserved spin-2 current (not the canonical stress tensor)
defining positive central charges $c$ and $a$. 
I calculate the values of $c$ and $a$ and study the
operator-product structure. Higher-spin
conformal spinors have no gauge invariance, admit
a standard definition of $c$ and $a$ and
can be coupled to Abelian and non-Abelian gauge fields in a renormalizable
way. At the quantum level, they contribute to the one-loop beta function
with the same sign as ordinary matter, admit a conformal window and
non-trivial interacting fixed points. There are composite operators of high
spin and low dimension, which violate the Ferrara--Gatto--Grillo theorem.
Finally, other theories, such as conformal antisymmetric tensors,
exhibit more severe internal problems.
This research is motivated by the idea that 
fundamental quantum field theories should be renormalization-group
(RG) interpolations between ultraviolet and infrared conformal fixed points,
and quantum irreversibility should be a
general principle of nature.}
\end{quote}

\vskip 4truecm

\noindent December, 1999\, - \, CERN-TH/99-383

\vfill\eject  

\section{Statement of the problem}

In the approach to quantum field theory as a radiative interpolation between
pairs of conformal fixed points (see \cite{proc} for a brief survey)
important physical information is given by the values of the
central charges $c$ and $a$ at the fixed points
and their dependence on the energy scale, in particular the differences
$c_{\rm UV}-c_{\rm IR}$ and $a_{\rm UV}-a_{\rm IR}$.
In a conformal theory, $c$
and $a$ are defined by the trace anomaly computed in the presence
of an external gravitational field. The quantity $%
c $ multiplies the square of the Weyl tensor $W_{\mu \nu \rho \sigma }^{2}$
and is the coefficient of the two-point function of the stress tensor. The
quantity $a$ multiplies the Euler density ${\rm G}_{4}=\varepsilon _{\mu \nu
\rho \sigma }\varepsilon _{\alpha \beta \gamma \delta }R^{\mu \nu \alpha
\beta }R^{\rho \sigma \gamma \delta }$. A third term, $\Box R$, is
multiplied by a coefficient $a^{\prime }$: 
\begin{equation}
\Theta =\frac{1}{(4\pi )^{2}}\left[ -cW^{2}+\frac{a}{4}{\rm G}_{4}-\frac{2}{3%
}a^{\prime }\Box R\right] .  \label{1.1}
\end{equation} 
In various UV-free supersymmetric models it is possible to compute $c$ and $a$ 
exactly in the IR limit \cite{noi}. In classically conformal
renormalizable theories, a simple non-perturbative
 formula for the total RG flow of $a$
\cite{athm} holds, which shows the phenomenon of 
quantum irreversibility
(the inequality $a_{\rm UV}\geq a_{\rm IR}$) \cite{athm,at6d,cea}.
The quantity $a$ is interpreted as a
counter of the massless degrees of freedom.

A natural problem is to study the generalization of these results
to higher-spin fields,
gravity in particular.
It is not straightforward to define 
the central charges $c$ and $a$ for gravity 
or higher-spin fields, because they are not conformal.
The trace anomaly for higher-spin fields was studied 
by Christensen and Duff \cite{duff} long ago and
contains the square of the Ricci curvature $R:$%
\[
\Theta =\frac{1}{(4\pi )^{2}}\left[ -cW^{2}+\frac{a}{4}{\rm G}_{4}+\zeta
R^{2}-\frac{2}{3}a^{\prime }\Box R\right] ,
\]
a sign that the field is not conformal. The definition of $c$ and $a
$ from the trace anomaly is unambiguous only if there is no such term. 
 Explicitly, for spin 3/2 and spin 2 we find, from table II of ref. 
\cite{duff} (omitting the $\Box R$-term): 
\begin{eqnarray*}
\Theta _{3/2} &=&\frac{1}{360(4\pi )^{2}}\left[ 255\,W^{2}-22\,\frac{{\rm G}%
_{4}}{4}+{\frac{61}{2}}R^{2}\right] , \\
\Theta _{2} &=&\frac{1}{360(4\pi )^{2}}\left[ -297\,W^{2}-127\,\frac{{\rm G}%
_{4}}{4}-{\frac{717}{2}}R^{2}\right] .
\end{eqnarray*}
We see that the graviton and gravitino contributions to the $R^{2}$-term
have opposite signs. This means that a suitable combination of gravitons and
gravitinos can cancel the $R^{2}$-term and might define 
a good higher-spin generalization of conformal
field theory. It would be interesting to 
have a comprehensive list of the
theories with vanishing $R^{2}$-term, starting from the analysis of
Christensen and Duff, and understand if they can be considered conformal for
all purposes. The disappearance of the $R^{2}$-term is a necessary
condition, but might not be sufficient.
Observe, however, that it is not simple
to construct manifestly gauge-invariant stress tensors for 
higher-spin fields and the $%
R^{2}$-term might depend on the definition, the gauge-fixing
or the scheme choice. On the other hand,
the conformal higher-spin fields,
which are 
studied in this paper, have better properties in connection with
these issues.

If sensible definitions of $c$ and $a$ are not
separately available, it might be interesting to look for an appropriate
generalization of the  subclass of
conformal field theories that have $c=a$. These theories share
various properties with two-dimensional conformal field theory \cite{cea}.
We see that 
the difference $c-a$ multiplies the unique term containing the Riemann
tensor in the trace anomaly, $R_{\mu \nu \rho \sigma }^{2}$
in the trace anomaly. An appropriate
definition of $c=a$ theories of gravitons and gravitinos,
or higher-spin fields in general, might include
the theories whose trace anomaly contains only $R_{\mu \nu }^{2}$, $R^{2}$
and $\Box R$, but not $R_{\mu \nu \rho \sigma }^{2}$. 
In arbitrary even dimensions the trace anomaly of the $c=a$ theories
contains the ``minimal amount'' of Riemann tensors, as shown in ref. \cite
{cea}.
However, the $c=a$ theories, certainly mathematically interesting, appear to 
be phenomenologically disfavoured (sect. 1.2).

In general, when do we have sensible definitions of $c$ and $a$?
We first demand that the classical theory
be conformal, so that no $R^2$-term is present
in the trace anomaly. Secondly, we would like that $c$ and $a$
be both positive. Negative central charges
represent a severe violation of unitarity.
The purpose of this paper is to 
explore a large variety of theories, old and new, in this  spirit.
We begin with higher-derivative theories (section 2)
and show that they have negative $c$ and $a$. 
We continue by classifying the higher-spin conformal
theories. Fermions admit a straightforward coupling to gravity
and gauge fields, so that in this case our program
can be carried over to the end. There is evidence 
of a conformal window and that 
these theories obey the irreversibility property.
On the other hand, higher-spin conformal bosons do not
admit a straightforward coupling to gravity. Nevertheless, I show
that there exists a suitable spin-2 current that has all the features
to define appropriate $c$ and $a$. I compute their values in a simple case and 
show that they are positive.

Higher-spin conformal field theories
 are not unitary \cite{vannieu}, but have a
number of interesting features (of which conformal invariance is just the
most important), which makes them interesting either as a laboratory for
investigations in the spirit of \cite{proc} and the questions raised above,
or for the description of physical phenomena in limited energy ranges.
In some cases, they have a
positive-definite action in the Euclidean framework. 
Symmetric conformal
tensors have a higher-derivative gauge symmetry, which is investigated
in detail. It is the unique gauge
transformation compatible with the conformal symmetry. Moreover, these
theories admit proper definitions of field strengths, dual field strengths,
Chern--Simons forms, topological invariants, etc. 

The non-unitarity of these theories
can show up in the negative sign of $c$ or $a$, as remarked above.
But even when 
$c$ and $a$ are positive, there might be effects on the anomalous 
dimensions of the operators generated by the OPE of two stress tensors.
For example, the
Ferrara--Gatto--Grillo theorem \cite{grillo} states that, in a unitary
theory, primary composite
operators with spin $s$ should have a total dimension greater than or equal
to $2+s$. This theorem is here manifestly violated.
Indeed, the higher-derivative gauge invariance allows for
``multiply-conserved'' currents with dimensions $\Delta =2+s,$ $1+s,\ldots $%
, $3$. Some of these operators will be constructed explicitly.

There have been earlier works on conformal field equations of spin-2 \cite
{bayen,drew,aragone,barut,delfino} and spin-3/2 \cite{barut} fields. These
theories are particular cases of the ones presented below. To my knowledge
the relationship between conformal invariance and higher-derivative gauge
invariance was not known. Recently, related theories have attracted some
interest in the domain of nuclear physics, where the purpose is to account
for the hadronic resonances, such as the spin-3/2 $\Delta $(1232) \cite
{haberzett,belgi}. I believe that the properties outlined here might be
useful in this domain, at least in a definite energy range.
The hope is that dynamical effects might make these
theories consistent at low energies,
thanks to quantum irreversibility itself (the ghosts might disappear above
the Planck length, far before the physically observable degrees of freedom)
or to a generalization of the Nachtmann theorem \cite{nach}.

Finally, I remark
that the study of higher-spin conformal field theory is in some sense
complementary to the Fradkin--Vasiliev higher-spin field theory \cite
{vasiliev}, which is unitary, but not conformal.

The plan of the paper is as follows.  Before entering
the technical part, I explain in sect. 1.1 
the reasons why it is physically attractive
to investigate higher-spin
theories focusing on $c$ and 
$a$ and the irreversibility.
Section 2 contains observations about
higher-derivatives conformal field theories and the rest
of the paper is about higher-spin
conformal field theories. 
I present the bosonic conformal fields
in section 3, the fermionic fields in section 5. Section 4 is devoted to a
detailed analysis of the spin-2 field, with computations of $c$ and $a$ and
a study of the operator-product (OPE) structure. In section 5 the
contribution to the gauge beta function from conformal spin-3/2 matter
fermions is computed. I work in the Euclidean
framework throughout this paper.

\subsection{Other motivations for the approach suggested here}

The search for a consistent
formulation of higher-spin fields is difficult, and
it might help to establish some
general guidelines. This section is devoted to explaining
why the properties of
the central charges $c$ and $a$ and quantum irreversibility
might be good for this.
The ideas contained here should be
meant as a work hypothesis. They apply to renormalizable, unitary theories
of fields with spin 0, 1/2 and 1, but their generalization
to higher-spin fields, gravity in particular,
is  not evident. Yet, it is a bit uncomfortable that
fields of spin 0, 1/2 and 1 have such 
a different status from fields of higher
spin and a ``unified''
description would be desirable.

The two  principles underlying
the proposed approach are: ({\it i}) 
a ``conformal hypothesis'', saying that 
{\it every quantum field theory describing the phenomena of
nature should be a renormali\-za\-ti\-on-group  interpolation between a
UV conformal field theory and an IR conformal field theory;}
and ({\it ii}) the irreversibility principle, stating that
{\it all fundamental theories of nature should
be quantum irreversible},
i.e. satisfy $a_{\rm UV}\geq a_{\rm IR}$. 

We can distinguish between a strong form of the conformal
hypothesis, when the classical action
 is conformal-invariant and 
all dimensionful parameters descend from 
the RG scale $\mu $, 
and a weak form of this hypothesis, 
when Newton's constant and, eventually, other
non-renormalizable parameters, descend from $\mu $, like $\Lambda _{%
{\rm QCD}}$, but the
classical action
is allowed to contain super-renormalizable terms
and masses and be conformal only in their absence. 
Massless QCD obeys the strong form of
the conformal hypothesis (the masses of hadrons are
proportional to $\Lambda _{{\rm QCD}}$ and therefore descend from $\mu $),
while
massive QCD obeys the weak form of the 
conformal hypothesis (the pion
masses and corrections to hadron masses are generated by the
quark masses, i.e. they do not descend from $\mu$).

Theories obeying the strong conformal hypothesis fall under the treatment
of refs. \cite{athm,at6d} and admit
a general formula expressing the difference $a_{{\rm UV}}-a_{{\rm %
IR}}$ \cite{athm} as the invariant area of the graph of the beta function.
Theories obeying the weak conformal hypothesis
 also  satisfy the
inequality $a_{{\rm UV}}\geq a_{{\rm IR}}$\footnotemark\footnotetext{
A mass term ${\frac{1}{2}}m^{2}\varphi ^{2}$ or, in general,
super-renormalizable terms, has the effect of killing degrees of freedom in
the IR, while the UV is left unchanged \cite{appel}. 
Roughly, this happens because the limit $%
m\rightarrow \infty $ is compatible only with $\varphi \equiv 0$.}, 
but the actual value of $a_{{\rm %
UV}}-a_{{\rm IR}}$ is not measured
by  the same formula \cite{cea}. These two cases can be
assimilated to each other at the quantitative level when $c=a$ \cite
{cea}. It is not clear at present how to link these two forms
of irreversibility in the general case $c\neq a$, 
but presumably there is no 
need to either, since the effects of masses can usually be 
included straightforwardly \cite{noi2}.
Observe that the irreversibility property implies that
a perturbative formulation of quantum field theory is meaningful only from the
UV. Since degrees of freedom are lost
from the UV to the IR, a quantum field
theory formulated from the IR (such as in IR-free theories) should be plagued
by inconsistencies or be trivial\footnotemark\footnotetext{
For example, the
Landau poles in QED, the non-perturbative triviality
of the $\lambda \phi ^{4}$-theory, the perturbative
non-renormalizability of quantum  gravity. 
There is a good amount of evidence
that QCD, instead, is fully consistent.}.

If the ultimate theory of the world falls into one of the two classes
mentioned here,
then the 
theories not obeying the (strong or weak) conformal hypothesis should be
viewed as low-energy effective theories descending from high-energy
fundamental theories that obey the conformal hypothesis.
In particular, any non-renormalizable interaction, and
gravity in particular, should be a
low-energy effect.
Observe that, if it were not so, non-renormalizable terms,
admitting that we can make sense out of them without additional
fields in the theory,
are expected to violate the irreversibility principle.
For example, a term $\varphi ^{6}/m^{2}$ forces the field $\varphi$
to vanish identically 
in the UV\ limit $m\rightarrow 0$ and leaves the IR unchanged. 
In certain supersymmetric theories \cite{noi2} where
some treatment of non-renormalizable terms has been
claimed to be meaningful in
the context of the so-called electric--magnetic duality, it has been found
that non-renormalizable terms do violate the irreversibility
statement (see section 6 of \cite{noi2}).

Finally, a special place is occupied by the ``$c=a$ flows'',
i.e. flows that connect UV and IR fixed points in such a way that the
difference $c-a$ remains constant (not necessarily zero), or
at least $a_{%
{\rm UV}}-a_{{\rm IR}}=c_{{\rm UV}}-c_{{\rm IR}}$. The fixed points might or
not have $c=a$. For example, taking a direct product between a $c=a$ flow
connecting two $c=a$ fixed points and a free-field theory, we can obtain a $%
c=a$ flow connecting $c\neq a$ fixed points. These flows 
 also fall under the treatment
of \cite{cea}. In such cases dimensionless parameters can be assimilated to
dimensionful coupling constants, for example masses and, conceivably, also
the Newton constant, even if they are not induced by $\mu$. 
The weak point of these theories is that they
are phenomenologically disfavoured.
A necessary condition fot a $c=a$ flow is obtained by
comparing the values of $c$ and $a$ at energies admitting (approximate)
free-field descriptions. Using the free-field values
$c=
(N_{s}+6N_{f}+12N_{v})/120$ and  $a=(N_{s}+11N_{f}+62N_{v})/360$,
$N_{s,f,v}$ being the numbers of real scalars, Dirac fermions and vectors,
respectively,
the differences between the numbers of spin-0, 1/2,
1 fields at two such energies are related in this case by the formula 
\begin{equation}
2\Delta N_{s}+7\Delta N_{f}=26 \Delta N_{v}.  \label{gui}
\end{equation}
Comparing the UV and IR limits of massless QCD, we find $\Delta
N_{s}=-n_f^2+1$, $\Delta N_f= N_c n_f$, $\Delta N_v=N_c^2 -1$, where $N_c$
is the number of colours and $n_f$ is the number of quark flavours. It is
easy to check that the condition has no solution.
Similarly, the spectra of the known low-energy physics do not appear to obey
(\ref{gui}). For example, in the IR we can neglect the electron, but we have
to include it at energies comparable with its mass. Formula (\ref{gui})
implies that as soon as the electron, or other fermions, becomes important,
also vector fields should appear. There is no evidence of such a behaviour in
nature. Finally, neither the Standard Model nor QCD 
have $c=a$\footnotemark\footnotetext{We can check it in the
free-field limits. QED has $N_{v}=1,$ $N_{f}=1$ and
$c-a=-37/360$. Massless QCD\ has $N_{v}=8,$ $%
N_{f}=18$ and $c-a=-
41/180$. The electroweak theory has $N_{v}=4,$ $N_{f}=9/2,$ $%
N_{s}=4$ and $c-a=-43/180$. The Standard Model 
has $c-a=-293/720$. We see that $c-a$ is
always negative, which means that there are many vector fields.}.

We have analysed various 
phenomena of ordinary theories that might suggest valid guidelines for
possible generalizations to new theories and gravity.
These are, in summary, the 
reasons why I think that it is interesting to investigate
higher-derivative theories and higher-spin conformal field
theories starting from the properties
of $c$ and $a$. We are now ready for the more
technical part. We first study higher-derivative theories
and then higher-spin conformal theories.

\section{Higher-derivative conformal field theories}

I begin by studying $c$ and $a$ in higher-derivative conformal field theories.
The free higher-derivative scalar field is
interesting because it corresponds to the induced action for the conformal
factor $\phi $ in an ordinary 
renormalizable theory and is described by the lagrangian 
\begin{equation}
S=\frac{1}{2}\int {\rm d}^{4}x\sqrt{g}\left[ \phi \Delta _{4}\phi +\frac{Q}{%
16\pi }{\rm \tilde{G}}_{4}\phi \right] ,
\label{oppp}
\end{equation}
where $\Delta _{4}=\Box ^{2}+2R^{\mu \nu }\nabla _{\mu }\nabla _{\nu }-\frac{%
2}{3}R\Box +\frac{1}{3}(\nabla ^{\mu }R)\nabla _{\mu }$ \cite
{riegert,mottola} and ${\rm \tilde{G}}_{4}={\rm G}_{4}-\frac{8}{3}\Box R$ is
the ``pondered'' Euler density \cite{at6d}. Here $Q$ is a background charge.
It does affect the induced effective 
action for the gravitational field, but its contribution
is not of anomalous origin (see below). I keep $Q$ non-zero to show that it
cannot be used to make either $c$ or $a$ positive.

The theory (\ref{oppp})
has been studied in refs. \cite{mottola,odintsov} and is
the four-dimensional analogue of the free two-dimensional scalar field.
Non-unitarity is evident from the fact that $c$ and $a$ are negative: 
\begin{equation}
c=-\frac{1}{15},\qquad a=-\frac{7}{90}-Q^{2}.  \label{hcar}
\end{equation}
The values at $Q=0$ can be read from \cite{riegert,mottola} and are the
contributions coming from the determinant of $\Delta _{4}$. The
contribution of the background charge is found after a translation in the
functional integral and the integration over $\phi $:
\begin{eqnarray*}
S &=&\frac{1}{2}\int {\rm d}^{4}x\sqrt{g}\left( \phi +\frac{Q}{32\pi }{\rm 
\tilde{G}}_{4}\frac{1}{\Delta _{4}}\right) \Delta _{4}\left( \phi +\frac{Q}{%
32\pi }{\rm \tilde{G}}_{4}\frac{1}{\Delta _{4}}\right) -\frac{Q^{2}}{(4\pi
)^{2}}{\rm \tilde{G}}_{4}\frac{1}{\Delta _{4}}{\rm \tilde{G}}_{4} \\
&\rightarrow &\frac{1}{2}{\rm tr}\ln \left[ \sqrt{g}\Delta _{4}\right] -%
\frac{Q^{2}}{128(4\pi )^{2}}{\rm \tilde{G}}_{4}\frac{1}{\Delta _{4}}{\rm 
\tilde{G}}_{4}.
\end{eqnarray*}
The term proportional to $Q^{2}$ affects the quantity
$a$ by
the shift written in (\ref{hcar}). This can be seen using for example the
formulas of \cite{riegert}. 

No real value of $Q$ can give a positive $a$, as promised.
Moreover, the background charge has no effect on $c$. The negative value of $%
c$ can be checked by computing the stress-tensor two-point function and does
not depend on $Q$. The stress tensor reads 
\begin{eqnarray*}
T_{\mu \nu } &=&-\partial _{\mu }\Box \phi \partial _{\nu }\phi -\partial
_{\nu }\Box \phi \partial _{\mu }\phi -\frac{4}{3}\partial _{\mu }\partial
_{\alpha }\phi \partial _{\nu }\partial _{\alpha }\phi +\frac{2}{3}\partial
_{\mu }\partial _{\nu }\partial _{\alpha }\phi \partial _{\alpha }\phi
+2\Box \phi \partial _{\mu }\partial _{\nu }\phi  \\
&&+\delta _{\mu \nu }\left[ \frac{1}{3}\partial _{\alpha }\Box \phi \partial
_{\alpha }\phi +\frac{1}{3}\left( \partial _{\alpha }\partial _{\beta }\phi
\right) ^{2}-\frac{1}{2}\left( \Box \phi \right) ^{2}\right] +\frac{Q}{6\pi }%
\left( \partial _{\mu }\partial _{\nu }-\delta _{\mu \nu }\Box \right) \Box
\phi .
\end{eqnarray*}
The two-point function is 
\[
\langle T_{\mu \nu }(x)~T_{\rho \sigma }(0)\rangle =-\frac{1}{120\pi ^{4}}{%
\prod }_{\mu \nu ,\rho \sigma }^{(2)}\left( \frac{1}{|x|^{4}}\right) ,
\]
in agreement with the value of $c$ given above. We see that the
non-unitarity of the theory is exhibited by a violation of reflection
positivity. Similarly, the non-unitarity of non-conformal higher-derivative
theories, such as a scalar field with lagrangian ${\cal L}={\frac{1}{2}}\Box
\phi (\Box +m^{2})\phi $, is exhibited by poles with negative residues in
the propagator \cite{pais}.

A  way to change the signs of both central charges it to consider
``higher-derivative anticommuting scalar fields'', $\theta ,~\bar{\theta}.~$%
In this case $Q=0$ and 
\[
S=\int {\rm d}^{4}x\sqrt{g}\bar{\theta}\Delta _{4}\theta ,\qquad c=\frac{2}{%
15},\qquad a=\frac{7}{45}. 
\]
This theory can be coupled to, say, the electromagnetic field. In a flat
gravitational background the most general renormalizable lagrangian has a
finite number of parameters due to the statistics of $\theta :$%
\[
{\cal L}=\frac{1}{4}F_{\mu \nu }^{2}+|D_{\mu }D_{\mu }\theta |^{2}+iF_{\mu
\nu }\overline{D_{\mu }\theta }D_{\nu }\theta +|D_{\mu }\theta D_{\nu
}\theta |^{2}+\cdots , 
\]
where $D^{2}=D_{\mu }D_{\mu }$. Each term can be further multiplied by a
polynomial $1+h\bar{\theta}\theta $. Some simplification comes from the
invariance of the theory under the renormalizable change of variables 
\[
A_{\mu }\rightarrow A_{\mu }+i\alpha \bar{\theta}\overleftrightarrow{%
\partial _{\mu }}\theta , 
\]
$\alpha $ being a parameter of no physical interest.

The change of the statistics of the fields does not eliminate the
non-unitarity of the theory. Indeed, the low-dimensionality of $\theta ,~%
\bar{\theta}$ allows us to construct many operators violating the
Ferrara--Gatto--Grillo theorem. There are also operators satisfying
reflection positivity before the change of statistics and violating it
afterwards. For example 
\[
\langle (\partial _{\mu }\bar{\theta}\partial _{\mu }\theta )(x)\,(\partial
_{\nu }\bar{\theta}\partial _{\nu }\theta )(0)\rangle =-(\partial _{\mu
}\partial _{\nu }\langle \bar{\theta}(x)\,\theta (0)\rangle )^{2}<0. 
\]
Finally, there are also operators that have a vanishing two-point function,
such as two terms of the electromagnetic current: 
\[
j_{\mu }=i\left( \bar{\theta}\Box {\partial _{\mu }}\theta -\partial _{\mu
}\Box \bar{\theta}\theta +{\frac{1}{3}}\partial _{\alpha }\bar{\theta}%
\overleftrightarrow{\partial _{\mu }}\partial _{\alpha }\theta +{\frac{4}{3}}%
\Box \bar{\theta}\partial _{\mu }\theta -{\frac{4}{3}}\partial _{\mu }\bar{%
\theta}\Box \theta \right) ,\qquad j_{\mu }^{\prime }=-{\frac{i}{2}}\pi
_{\mu \alpha }\left( \bar{\theta}\overleftrightarrow{\partial _{\alpha }}%
\theta \right) . 
\]
We find, defining $J_{\mu }=aj_{\mu }+bj_{\mu }^{\prime }$, 
\[
\langle j_{\mu }(x)\,j_{\nu }(0)\rangle =\langle j_{\mu }^{\prime
}(x)\,j_{\nu }^{\prime }(0)\rangle =0,\qquad \langle J_{\mu }(x)\,J_{\nu
}(0)\rangle =-{\frac{ab}{4\pi ^{2}}}\pi _{\mu \nu }\left( {\frac{1}{|x|^{4}}}%
\right) . 
\]

Despite the unitarity problem, renormalization of this theory is
well-behaved and very presumably there is a conformal window, at least when
the gauge field is non-Abelian. Theories such as these are a useful laboratory
for the approach of \cite{proc}.

For fermionic theories 
\[
{\cal L}=\bar{\psi}\partial \!\!\!\slash \Box \psi ,
\]
we have found the stress tensor 
\begin{eqnarray}
T_{\mu \nu } &=&h\left\{ \bar{\psi}(\gamma _{\mu }\partial _{\nu }+\gamma
_{\nu }\partial _{\mu })\Box \psi -\Box \bar{\psi}(\gamma _{\mu }%
\overleftarrow{\partial _{\nu }}+\gamma _{\nu }\overleftarrow{\partial _{\mu
}})\psi +3\Box \bar{\psi}(\gamma _{\mu }\partial _{\nu }+\gamma _{\nu
}\partial _{\mu })\psi \right.   \nonumber \\
&&-3\bar{\psi}(\gamma _{\mu }\overleftarrow{\partial _{\nu }}+\gamma _{\nu }%
\overleftarrow{\partial _{\mu }})\Box \psi -{\frac{2}{3}}(\partial _{\mu }%
\bar{\psi}\overleftrightarrow{\partial \!\!\!\slash}\partial _{\nu }\psi
+\partial _{\nu }\bar{\psi}\overleftrightarrow{\partial \!\!\!\slash}%
\partial _{\mu })\psi +2\partial _{\alpha }\bar{\psi}(\gamma _{\mu }%
\overleftrightarrow{\partial _{\nu }}+\gamma _{\nu }\overleftrightarrow{%
\partial _{\mu }})\partial _{\alpha }\psi   \nonumber \\
&&-{\frac{10}{3}}(\bar{\psi}\overleftarrow{\partial \!\!\!\slash}\partial
_{\mu }\partial _{\nu }\psi -\partial _{\mu }\partial _{\nu }\bar{\psi}%
\partial \!\!\!\slash\psi )-{\frac{2}{3}}(\bar{\psi}\partial \!\!\!\slash %
\partial _{\mu }\partial _{\nu }\psi -\partial _{\mu }\partial _{\nu }\bar{%
\psi}\overleftarrow{\partial \!\!\!\slash}\psi )  \nonumber \\
&&\left. +{\frac{1}{3}}\delta _{\mu \nu }[7(\bar{\psi}\overleftarrow{%
\partial \!\!\!\slash}\Box \psi -\Box \bar{\psi}\partial \!\!\!\slash\psi
)-2\partial _{\alpha }\bar{\psi}\overleftrightarrow{\partial \!\!\!\slash}%
\partial _{\alpha }\psi ]\right\} ,
\end{eqnarray}
by imposing conservation and tracelessness. It is not straightforward to fix
the overall factor $h$ from the coupling to gravity (a
Weyl-invariant coupling to external gravity might not exist). The factor
could be fixed unambiguously with the OPE technique of sect. 4 or the
Noether method, but here we do not need it, since our primary concern is to
show that $c$ is negative, independently of the value of $h$. We find 
\[
\langle T_{\mu \nu }(x)\,T_{\rho \sigma }(0)\rangle =-{\frac{8h^{2}}{15\pi
^{4}}}{\prod }_{\mu \nu ,\rho \sigma }^{(2)}\left( {\frac{1}{|x|^{4}}}%
\right) <0.
\]
We might wonder whether the situation changes in higher dimensions, but it
is not so. I have checked that a free scalar field with action $\frac{1}{2}%
(\Box \phi )^{2}$ in six dimensions has, again, $c<0$. The stress tensor
reads 
\begin{eqnarray*}
T_{\mu \nu } &=&h\left\{ \frac{3}{4}\partial _{\mu }\partial _{\alpha }\phi
\partial _{\nu }\partial _{\alpha }\phi -\frac{3}{2}\Box \phi \partial _{\mu
}\partial _{\nu }\phi +\partial _{\nu }\Box \phi \partial _{\mu }\phi
+\partial _{\mu }\Box \phi \partial _{\nu }\phi -\frac{1}{2}\partial _{\mu
}\partial _{\nu }\partial _{\alpha }\phi \partial _{\alpha }\phi \right.  \\
&&-\frac{1}{4}\phi \Box \partial _{\mu }\partial _{\nu }\phi +\left. \delta
_{\mu \nu }\left[ -\frac{1}{4}\partial _{\alpha }\Box \phi \partial _{\alpha
}\phi -\frac{1}{8}\left( \partial _{\alpha }\partial _{\beta }\phi \right)
^{2}+\frac{1}{4}\left( \Box \phi \right) ^{2}\right] \right\} 
\end{eqnarray*}
and the two-point function is 
\[
\langle T_{\mu \nu }(x)\,T_{\rho \sigma }(0)\rangle =-{\frac{25h^{2}}{%
86016\pi ^{6}}}{\prod }_{\mu \nu ,\rho \sigma }^{(2)}\left( {\frac{1}{|x|^{6}%
}}\right) <0.
\]

Summarizing, higher-derivative theories with ordinary statistics
have negative
central char\-ges, which means that the ghost
degrees of freedom prevail over the physical ones. 
Yet, these theories can be conformal at the classical level
and renormalizable at the quantum level. While the unitarity violations can
hardly be removed completely, it might be possible that in some conformal
theories certain problems are less serious, so that $c$ and $a$ be positive,
despite of the underlying non-unitarity. Higher-spin conformal field theories,
which I analyse in the rest of the paper, appear to have this property.

\section{Conformal bosonic fields}

The simplest example of higher-spin conformal field theory is the free
spin-2 conformal field.

Let $\chi _{\mu \nu }$ be symmetric and traceless. The action 
\begin{equation}
S=\int {\cal L}_{1}=\int \frac{1}{2}(\partial _{\mu }\chi _{\nu \rho })^{2}-%
\frac{2}{3}(\partial _{\mu }\chi _{\mu \nu })^{2}  \label{action1}
\end{equation}
is invariant with respect to the 
coordinate inversion $x^{\mu }\rightarrow \frac{%
x^{\mu }}{|x|^{2}}$. Under this transformation the tensor $\chi _{\mu \nu }$
transforms as 
\[
\chi _{\mu \nu }(x)\rightarrow |x|^{2}I_{\mu \rho }^{{}}(x)I_{\nu \sigma
}^{{}}(x)\chi _{\rho \sigma }(x), 
\]
where $I_{\mu \nu }^{{}}(x)=\delta _{\mu \nu }^{{}}-2x_{\mu }x_{\nu
}/|x|^{2} $. This invariance fixes uniquely the action $S$, and the
lagrangian ${\cal L}_{1}$ up to total derivatives. A better choice of the
total derivatives leads to the lagrangian 
\begin{equation}
{\cal L}=\frac{1}{2}(\partial _{\mu }\chi _{\nu \rho })^{2}-\frac{1}{2}%
\partial _{\mu }\chi _{\nu \rho }\partial _{\nu }\chi _{\mu \rho }-\frac{1}{6%
}(\partial _{\mu }\chi _{\mu \nu })^{2}.  \label{action2}
\end{equation}
${\cal L}$ transforms as a scalar under coordinate inversion, ${\cal %
L\rightarrow }|x|^{8}{\cal L}$. The action $S$ is invariant under the
higher-derivative gauge transformation 
\begin{equation}
\delta \chi _{\mu \nu }=\partial _{\mu }\partial _{\nu }\Lambda -\frac{1}{4}%
\delta _{\mu \nu }\Box \Lambda ,  \label{invariance}
\end{equation}
but not with respect to the diffeomorphism-type transformation $\delta \chi
_{\mu \nu }=\partial _{\mu }\xi _{\nu }+\partial _{\nu }\xi _{\mu }$. The
lagrangian ${\cal L}$ is also invariant, while ${\cal L}_{1}$ is invariant
up to a total derivative.

The gauge transformation (\ref{invariance}) is compatible with the conformal
symmetry. This can be proved by observing that $\Lambda $ has dimension $-1$
and thus, under coordinate inversion, $\delta \chi _{\mu \nu }$ transforms
in the same way as $\chi _{\mu \nu }$: 
\[
\Lambda \rightarrow |x|^{-2}\Lambda ,\qquad \partial _{\mu }\partial _{\nu
}\Lambda -\frac{1}{4}\delta _{\mu \nu }\Box \Lambda \rightarrow
|x|^{2}I_{\mu \rho }^{{}}(x)I_{\nu \sigma }^{{}}(x)\left( \partial _{\rho
}\partial _{\sigma }\Lambda -\frac{1}{4}\delta _{\rho \sigma }\Box \Lambda
\right) . 
\]
To check this, observe that the derivative operator transforms as a vector
of dimension 1: $\partial _{\mu }\rightarrow |x|^{2}I_{\mu \nu
}^{{}}(x)\partial _{\nu }$.

The field equations read 
\[
\Box \chi _{\mu \nu }=\frac{2}{3}\partial _{\rho }\left( \partial _{\mu
}\chi _{\nu \rho }+\partial _{\nu }\chi _{\mu \rho }\right) -{\frac{1}{3}}%
\delta _{\mu \nu }\partial _{\rho }\partial _{\sigma }\chi _{\rho \sigma }. 
\]
Defining the vector field ${\cal A}_{\mu }=\partial _{\nu }\chi _{\mu \nu }$
and its field strength ${\cal F}_{\mu \nu }=\partial _{\mu }{\cal A}_{\nu
}-\partial _{\nu }{\cal A}_{\mu }$, the field equations and gauge invariance
imply 
\[
\partial _{\mu }{\cal F}_{\mu \nu }=0,\qquad \qquad \delta {\cal A}_{\mu }={%
\frac{3}{4}}\partial _{\mu }\Box \Lambda . 
\]

\subsection{Field strength}

The gauge symmetry (\ref{invariance}) leads to the introduction of a natural
field strength, 
\[
F_{\mu \nu \alpha }=\partial _{\mu }\chi _{\nu \alpha }-\partial _{\nu }\chi
_{\mu \alpha }-\frac{1}{3}\delta _{\mu \alpha }\partial _{\rho }\chi _{\rho
\nu }+\frac{1}{3}\delta _{\nu \alpha }\partial _{\rho }\chi _{\rho \mu }, 
\]
which is easily proved to be gauge-invariant. This field strength satisfies
a number of noticeable properties. First of all, we have the identities 
\begin{equation}
F_{\mu \nu \alpha }=-F_{\nu \mu \alpha },\qquad F_{\mu \nu \mu }=0,\qquad
F_{\mu \nu \alpha }+F_{\alpha \mu \nu }+F_{\nu \alpha \mu }=0.  \label{sym}
\end{equation}
The third identity will be called the {\it cyclic identity}. Secondly, the
lagrangian (\ref{action2}) can be written as 
\[
{\cal L}=\frac{1}{4}(F_{\mu \nu \alpha })^{2}, 
\]
which implies, in particular, that it is positive-definite, a fact that was
not evident from (\ref{action1}) and (\ref{action2}). Since ${\cal L}$ is a
conformal field of dimension 4, it is evident that the field strength is
itself a conformal field of dimension 2 and transforms as 
\[
F_{\mu \nu \alpha }\rightarrow |x|^{4}I_{\mu \rho }^{{}}(x)I_{\nu \sigma
}^{{}}(x)I_{\alpha \beta }^{{}}(x)F_{\rho \sigma \beta } 
\]
under coordinate inversion. The field equations read 
\begin{equation}
\partial _{\mu }F_{\mu \nu \alpha }+\partial _{\mu }F_{\mu \alpha \nu }=0.
\label{feq}
\end{equation}

It is convenient to introduce the dual of the field strength, as well as
self-dual and anti-self-dual field strengths: 
\[
\tilde{F}_{\mu \nu \alpha }=\frac{1}{2}\varepsilon _{\mu \nu \rho \sigma
}F_{\rho \sigma \alpha },\qquad F_{\mu \nu \alpha }^{\pm }=\frac{1}{2}\left(
F_{\mu \nu \alpha }\pm \tilde{F}_{\mu \nu \alpha }\right) . 
\]
Each of these tensors satisfies the same symmetry identities (\ref{sym}) as $%
F_{\mu \nu \alpha }$. We can also derive the ``Bianchi identity'' 
\[
\partial _{\mu }\tilde{F}_{\mu \nu \alpha }+\partial _{\mu }\tilde{F}_{\mu
\alpha \nu }=0. 
\]

There is a natural topological invariant and a ``Chern--Simons'' form: 
\[
F_{\mu \nu \alpha}\tilde{F}_{\mu \nu \alpha}=\partial _{\mu }\left(
\varepsilon _{\mu \nu \rho \sigma }\chi _{\nu \alpha}F_{\rho \sigma
\alpha}\right) . 
\]

Non-trivial interactions for conformal higher-spin theories can be
constructed, as power series in the field strength: 
\begin{equation}
{\cal L}=\frac{1}{4}\left( F_{\mu \nu \alpha }\right) ^{2}+\frac{1}{\Lambda
^{4}}\left\{ a\left[ \left( F_{\mu \nu \alpha }\right) ^{2}\right]
^{2}+bF_{\mu \nu \alpha }F_{\nu \rho \alpha }F_{\rho \sigma \beta }F_{\sigma
\mu \beta }+cF_{\mu \nu \alpha }F_{\nu \rho \beta }F_{\rho \sigma \alpha
}F_{\sigma \mu \beta }\right\} +\cdots  \label{serf}
\end{equation}
$\Lambda $ being some mass scale and $a$, $b$, $c$ being dimensionless
parameters. These vertices are non-trivial because they do not vanish
when the field equations (\ref{feq}) are satisfied. Our interest, however, is
mostly to look for non-trivial renormalizable interactions, which preserve
conformality at the classical level. These are more difficult to construct,
but are fundamental for the conformal hypothesis stated in the introduction.
Certain renormalizable interactions will be studied in this paper (for
fermionic higher-spin conformal theories), but a complete classification
will not be given here.

The coupling to gravity is not straightforward and might not exist at all.
Simple attempts to impose the compatibility between the gauge symmetry (\ref
{invariance}) and gravity generate terms that cannot be reabsorbed.
Nevertheless, this does not forbid a correct definition of $c$ and $a$ (see
section 3).

\subsection{Propagator and ghosts}

In
this section I calculate the propagators and discuss a number of important
features. The quantization is performed
in the BRS approach and in the framework of the functional integral.

The most natural gauge-fixing is 
\begin{equation}
\partial _{\mu }\partial _{\nu }\chi _{\mu \nu }=\partial _{\mu }{\cal A}%
_{\mu }=0.  \label{-1}
\end{equation}
The gauge-fixed lagrangian becomes 
\[
{\cal L}_{1}=\frac{1}{2}(\partial _{\mu }\chi _{\nu \rho })^{2}-\frac{2}{3}%
(\partial _{\mu }\chi _{\mu \nu })^{2}+b\,\partial _{\mu }\partial _{\nu
}\chi _{\mu \nu }-\frac{3}{4}\overline{C}\Box ^{2}C 
\]
and the BRS transformation reads 
\[
s\chi _{\mu \nu }=\partial _{\mu }\partial _{\nu }C-\frac{1}{4}\delta _{\mu
\nu }\Box C,\quad sC=0,\quad s\overline{C}=b,\quad sb=0. 
\]
Defining the projectors 
\begin{eqnarray*}
P_{1\mu \nu ,\rho \sigma } &=&\frac{1}{2}\left( \delta _{\mu \rho }\delta
_{\nu \sigma }+\delta _{\mu \sigma }\delta _{\nu \rho }\right) -\frac{1}{4}%
\delta _{\mu \nu }\delta _{\rho \sigma }, \\
P_{2\mu \nu ,\rho \sigma } &=&\frac{1}{3}\left( \partial _{\mu }\partial
_{\rho }\delta _{\nu \sigma }+\partial _{\mu }\partial _{\sigma }\delta
_{\nu \rho }+\partial _{\nu }\partial _{\rho }\delta _{\mu \sigma }+\partial
_{\nu }\partial _{\sigma }\delta _{\mu \rho }-\partial _{\mu }\partial _{\nu
}\delta _{\rho \sigma }-\partial _{\rho }\partial _{\sigma }\delta _{\mu \nu
}+\frac{1}{4}\Box \delta _{\mu \nu }\delta _{\rho \sigma }\right) \frac{1}{%
\Box }, \\
\varpi _{\mu \nu } &=&\partial _{\mu }\partial _{\nu }\frac{1}{\Box }-\frac{1%
}{4}\delta _{\mu \nu },
\end{eqnarray*}
we have the relations 
\[
P_{1}^{2}=P_{1},\quad P_{1}P_{2}=P_{2},\quad P_{2}^{2}=\frac{2}{3}P_{2}+%
\frac{4}{9}\varpi \varpi ,\quad P_{1}\varpi =\varpi ,\quad P_{2}\varpi
=\varpi ,\quad {\rm tr}\varpi \varpi =\frac{3}{4}. 
\]
The lagrangian, written as 
\[
{\cal L}_{1}+\frac{3}{4}\overline{C}\Box ^{2}C=-\frac{1}{2}\left( 
\begin{array}{cc}
\chi & b
\end{array}
\right) Q\left( 
\begin{array}{c}
\chi \\ 
b
\end{array}
\right) =-\frac{1}{2}\left( 
\begin{array}{cc}
\chi & b
\end{array}
\right) \Box \left[ 
\begin{array}{cc}
P_{1}-P_{2} & \quad -\varpi \\ 
-\varpi & \quad 0
\end{array}
\right] \left( 
\begin{array}{c}
\chi \\ 
b
\end{array}
\right) , 
\]
can be easily inverted to find the propagators, which are 
\[
\left\langle \left( 
\begin{array}{c}
\chi \\ 
b
\end{array}
\right) \left( 
\begin{array}{cc}
\chi & b
\end{array}
\right) \right\rangle =-\left[ 
\begin{array}{cc}
P_{1}+3P_{2}-\frac{16}{3}\varpi \varpi & \quad -\frac{4}{3}\varpi \\ 
-\frac{4}{3}\varpi & \quad 0
\end{array}
\right] \frac{1}{\Box }. 
\]
The $x$-space propagators can be written using 
\[
-\left( \frac{1}{\Box }\right) _{(x,0)}=\frac{1}{4\pi ^{2}}\frac{1}{|x|^{2}}%
,\qquad \frac{1}{\Box ^{2}}=-\frac{1}{16\pi ^{2}}\ln |x|^{2}\mu ^{2},\qquad 
\frac{1}{\Box ^{3}}=-\frac{|x|^{2}}{128\pi ^{2}}\left( \ln |x|^{2}\mu ^{2}-%
\frac{3}{2}\right) . 
\]

The field $b$ does not propagate, because $\langle b(x)\,b(0)\rangle
=\left\langle s\left( \overline{C}(x)\,b(0)\right) \right\rangle =0.$
Similarly, $\langle b(x)\,\chi _{\mu \nu }(0)\rangle $ vanishes on-shell,
i.e. when saturated by $\chi $-polarizations satisfying the gauge-fixing
condition $\partial _{\mu }\partial _{\nu }\chi _{\mu \nu }=0$.

The ghosts of the theory are the components 
${\cal A}_{\mu }$ and a way to single out the 
(two) physical degrees of freedom is to set
\begin{equation}
{\cal A}_\mu=\partial ^{\mu }\chi _{\mu \nu }=0.  \label{zero}
\end{equation}
This condition has no dynamical origin (the theory has
${\cal A}_\mu\neq 0$) and is here meant for a pedagogical purpose.
Observe that only when ${\cal A}_\mu=0$ 
the field equations reduce to an ordinary wave equation
for $\chi _{\mu \nu }$, $\Box \chi _{\mu \nu }=0.$ Moreover, the propagator,
saturated with $\chi $-polarizations, becomes  $|\chi _{\mu \nu
}(k)|^2/k^2$ in this case. The gauge-fixing condition $\partial
_{\mu }\partial _{\nu }\chi _{\mu \nu }=0$ gives, in momentum space, 
\begin{equation}
\chi _{00}+\hat{n}_{i}\hat{n}_{j}\chi _{ij}=2\hat{n}_{i}\chi _{0i},
\label{uno}
\end{equation}
where $\hat{n}_{i}=k_{i}/k_{0}$, $i=1,2,3$, and $k_{0}^{2}=k_{i}^{2}$. 
When ${\cal A}_\mu=0$, the
additional conditions $\partial _{\mu }\chi _{\mu i}=0$ give 
\[
\hat{n}_{j}\chi _{ij}=\chi _{0i}, 
\]
which, reinserted into (\ref{uno}), also give 
\[
\chi _{00}=\hat{n}_{i}\chi _{0i}, 
\]
i.e. $\partial _{\mu }\chi _{\mu 0}=0,$ justifying (\ref{zero}). The
condition of vanishing trace for $\chi_{\mu\nu}$ gives $\chi
_{ii}=-\chi_{00}=\hat{n}_{i}\hat{n}_{j}\chi _{ij}$. We have therefore 
\[
|\chi _{\mu \nu }|^{2}=|\chi _{ij}|^{2}+|\chi _{ii}|^{2}-2|\hat{n}_{j}\chi
_{ij}|^{2}. 
\]
Let us choose $\hat{n}_{i}=(0,0,1).$ The condition $\chi _{ii}=\hat{n}_{i}%
\hat{n}_{j}\chi _{ij}$ gives $\chi _{22}=-\chi _{11}$ and finally 
\[
|\chi _{\mu \nu }|^{2}=2(|\chi _{11}|^{2}+|\chi _{12}|^{2})\geq 0. 
\]
Concluding, 
the two physical degrees of freedom are $\chi _{11}$ and $\chi _{12}$,
the unphysical degrees of freedom are ${\cal A}_\mu$.
The question is which of the two prevail in $c$ and $a$.
If the central charges are positive, the physical degrees of freedom prevail
over the unphysical ones. We compute $c$ and $a$ in section 4.

\subsection{Arbitrary integer spin}

Let $\chi _{\mu _{1}\cdots \mu _{s}}$ be a completely symmetric and
completely traceless tensor. Invariance of the action under the
transformation 
\begin{equation}
\chi _{\mu _{1}\cdots \mu _{s}}\rightarrow |x|^{2}I_{\mu _{1}}^{\nu
_{1}}(x)\cdots I_{\mu _{s}}^{\nu _{s}}(x)\chi _{\nu _{1}\cdots \nu _{s}}
\label{confo}
\end{equation}
fixes uniquely the lagrangian 
\[
{\cal L}_{1}=\frac{1}{2}(\partial _{\alpha }\chi _{\mu _{1}\cdots \mu
_{s}})^{2}-\frac{s}{s+1}(\partial _{\alpha }\chi _{\alpha \mu _{2}\cdots \mu
_{s}})^{2}, 
\]
up to the overall factor and total derivatives. ${\cal L}_1$ reduces to the
usual vector lagrangian for $s=1$ and to the free real-scalar theory for $%
s=0.$ The action is invariant under the gauge transformation 
\[
\delta \chi _{\mu _{1}\cdots \mu _{s}}=\partial _{\mu _{1}}\cdots \partial
_{\mu _{s}}\Lambda -{\rm traces}, 
\]
which, as before, is compatible with (\ref{confo}), when taking into account
that $\Lambda $ has dimension $1-s$ and transforms as $\Lambda \rightarrow
|x|^{2(1-s)}\Lambda $ under coordinate inversion.

The field strength reads 
\[
F_{\mu \nu \mu _{2}\cdots \mu _{s}}=\partial _{\mu }\chi _{\nu \mu
_{2}\cdots \mu _{s}}-\partial _{\nu }\chi _{\mu \mu _{2}\cdots \mu _{s}}-%
\frac{1}{s+1}\sum_{i=2}^{s}\left( \delta _{\mu \mu _{i}}\partial _{\alpha
}\chi _{\alpha \nu \mu _{2}\cdots \widehat{\mu _{i}}\cdots \mu _{s}}-\delta
_{\nu \mu _{i}}\partial _{\alpha }\chi _{\alpha \mu \mu _{2}\cdots \widehat{%
\mu _{i}}\cdots \mu _{s}}\right) . 
\]
A hat denotes indices that have to be omitted. As before, the field strength
is gauge-invariant and conformal. It is completely symmetric in $\mu
_{2}\cdots \mu _{s}$ and antisymmetric in $\mu \nu .$ Furthermore, it is
completely traceless, not only in the indices $\mu _{2}\cdots \mu _{s}$, but
also with respect to the remaining contraction: 
\begin{equation}
F_{\mu \nu \nu \mu _{3}\cdots \mu _{s}}=0.  \label{trac}
\end{equation}
Finally, it satisfies the cyclic condition 
\begin{equation}
F_{\mu \nu \alpha \mu _{3}\cdots \mu _{s}}+F_{\alpha \mu \nu \mu _{3}\cdots
\mu _{s}}+F_{\nu \alpha \mu \mu _{3}\cdots \mu _{s}}=0.  \label{cyc}
\end{equation}
The conformal, positive-definite lagrangian can be written as 
\[
{\cal L}=\frac{1}{4}(F_{\mu \nu \mu _{2}\cdots \mu _{s}})^{2}. 
\]
The field equations and Bianchi identities are 
\[
\partial _{\mu }F_{\mu \alpha _{1}\cdots \alpha _{s}}+{\rm perms}(\alpha
_{1}\cdots \alpha _{s})=0,\qquad \quad \partial _{\mu }\tilde{F}_{\mu \alpha
_{1}\cdots \alpha _{s}}+{\rm perms}(\alpha _{1}\cdots \alpha _{s})=0. 
\]
Dual and self-dual field strengths are defined as 
\[
\tilde{F}_{\mu \nu\alpha _{2}\cdots \alpha _{s}}={\frac{1}{2}}%
\varepsilon_{\mu\nu\rho\sigma} {F}_{\rho\sigma \alpha_{2}\cdots \alpha
_{s}},\qquad {F}^\pm_{\mu \nu\alpha _{2}\cdots \alpha _{s}}= {\frac{1}{2}}%
\left( {F}_{\mu \nu\alpha _{2}\cdots \alpha _{s}} \pm \tilde {F}_{\mu
\nu\alpha _{2}\cdots \alpha _{s}}\right) 
\]
and satisfy the traceless and cyclic conditions (\ref{trac}) and (\ref{cyc}).

There is a topological invariant, proportional to the integral of 
\[
F_{\mu \nu \mu _{2}\cdots \mu _{s}}\tilde{F}_{\mu \nu \mu _{2}\cdots \mu
_{s}}=\partial _{\mu }\left( \varepsilon _{\mu \nu \rho \sigma }\chi _{\nu
\mu _{2}\cdots \mu _{s}}F_{\rho \sigma \mu _{2}\cdots \mu _{s}}\right) . 
\]
The equality can be proved by using the Bianchi identity and (\ref{trac}).

The stress tensor (see the discussion of sect. 4.1) is 
\[
T_{\mu \nu }={\rm const.}\, F_{\mu \alpha _{1}\cdots \alpha _{s}}^{+}F_{\nu
\alpha _{1}\cdots \alpha _{s}}^{-}. 
\]
Tracelessness is straightforward, while the proof of conservation follows
the same line as in the spin-2 case. The procedure to fix
the overall factor and the comparison with the Noether tensor are discussed in
detail for $s=2$. Higher-spin tensor currents can be constructed using the
recipes of \cite{high,OPE}.

\subsection{Implications of the higher-derivative gauge invariance on
correlators}

The general form of the two-point function of a conformal composite operator 
${\cal O}_{\mu _{1}\cdots \mu _{s}}$ with spin $s$ is, in the notation of 
\cite{high,OPE}: 
\begin{equation}
\langle {\cal O}_{\mu _{1}\cdots \mu _{s}}(x)~{\cal O}_{\nu _{1}\cdots \nu
_{s}}(0)\rangle =c_{s}\frac{1}{(|x|\mu )^{2h_{s}}}{\prod }_{\mu _{1}\cdots
\mu _{s},\nu _{1}\cdots \nu _{s}}^{(s)}\left( \frac{1}{|x|^{4}}\right) ,
\label{cor}
\end{equation}
where ${\prod }_{\mu _{1}\cdots \mu _{s},\nu _{1}\cdots \nu _{s}}^{(s)}$ is
the unique differential operator of degree $2s$ that is completely symmetric
and traceless in $\mu _{1}\cdots \mu _{s}$ and $\nu _{1}\cdots \nu _{s}$,
symmetric in the exchange $\mu \leftrightarrow \nu $, conserved with respect
to any index. For example, $\pi _{\mu \nu }=\partial _{\mu }\partial _{\nu
}-\delta _{\mu \nu }\Box $ for $s=1$, while ${\prod }_{\mu \nu ,\rho \sigma
}^{(2)}=\frac{1}{2}(\pi _{\mu \rho }\pi _{\nu \sigma }+\pi _{\mu \sigma }\pi
_{\nu \rho })-\frac{1}{3}\pi _{\mu \nu }\pi _{\rho \sigma }.$ The factor $%
c_{s}$ is a constant (higher-spin central charge) and $h_{s}$ is equal to $%
\delta _{s}-s-2,$ where $\delta _{s}$ is the total dimension of the operator 
${\cal O}_{\mu _{1}\cdots \mu _{s}}$.

If the operator ${\cal O}_{\mu _{1}\cdots \mu _{s}}$ couples to a conformal
higher-spin field $\chi _{\mu _{1}\cdots \mu _{s}}$, via a vertex ${\cal O}%
_{\mu _{1}\cdots \mu _{s}}\chi _{\mu _{1}\cdots \mu _{s}}$, then the
following ``multiple-conservation'' condition holds: 
\begin{equation}
\partial _{\mu _{1}}\cdots \partial _{\mu _{s}}{\cal O}_{\mu _{1}\cdots \mu
_{s}}=0.  \label{multicon}
\end{equation}
An ordinary conservation condition $\partial _{\mu _{s}}{\cal O}_{\mu
_{1}\cdots \mu _{s}}=0$ implies $h_{s}=0$. Instead, applying the
multiple-conservation condition (\ref{multicon}) to the correlator (\ref{cor}%
), we find that $h_{s}$ can take an arbitrary integer value between $0$ and $%
1-s.$ Consequently, we have the following spectrum of allowed dimensions: 
\begin{equation}
\delta _{s}=2+s,1+s,\cdots ,3.  \label{dime}
\end{equation}
Observe that only the operators of dimension 3 need $s$ divergences to be
annihilated. Operators of higher dimension are allowed to satisfy more
restrictive conditions. In particular, operators of dimension $2+s$ can be
conserved in the usual sense ($\partial _{\mu _{1}}{\cal O}_{\mu _{1}\cdots
\mu _{s}}=0$), operators of dimension $1+s$ can be annihilated by two
divergences ($\partial _{\mu _{1}}\partial _{\mu _{2}}{\cal O}_{\mu
_{1}\cdots \mu _{s}}=0$), etc.

The Ferrara--Gatto--Grillo theorem \cite{grillo} says that in a unitary
theory primary conformal operators with spin-$s$ have dimensions $\delta
_{s}\geq 2+s$. This property is here violated. We see from (\ref{dime}) that
the minimal allowed dimension is $3$. This feature is relevant to the
conformal hypothesis stated in the introduction: the interaction vertex 
\[
{\cal O}_{\mu _{1}\cdots \mu _{s}}\chi _{\mu _{1}\cdots \mu _{s}}
\]
is renormalizable if ${\cal O}_{\mu _{1}\cdots \mu _{s}}$ is such an
operator of dimension $3$; therefore, in our theories, renormalizable
higher-spin interactions are not ruled out in a trivial way
(see section 5.1).

\section{The spin-2 conformal boson in detail}

In this section I study the stress tensor of the spin-2 conformal boson,
compute its two-point function and OPE, and extract the values
of the central charges $c$
and $a$. The result is that both $c$ and $a$ have positive values. 

\subsection{Stress-tensor, spin-2 currents and the definitions of
$c$ and $a$}

The Noether method produces a non-gauge-invariant, non-symmetric, traceful
stress tensor 
\[
T_{\mu \nu }^{{\rm N}}=\partial _{\mu }\chi _{\alpha \beta }F_{\nu \alpha
\beta }-\frac{1}{4}\delta _{\mu \nu }F_{\alpha \beta \gamma }^{2}. 
\]
This operator is conserved in $\nu $ ($\partial _{\nu }T_{\mu \nu }=0$),
gauge-invariant and traceless up to total derivatives, and it does not
transform simply under coordinate inversion
$x_\mu\rightarrow x_\mu/|x|^2$. For this reason, it is not
easy to use the Noether tensor to extract $c$ and $a$. Moreover, no
improvement term $\psi _{\mu \nu \lambda }=-\psi _{\mu \lambda \nu
}$ appears to be such that $T_{\mu \nu }^{{\rm N}}+\partial _{\lambda }\psi
_{\mu \nu \lambda }$ transform correctly and be gauge-invariant.

There exists, nevertheless, a remarkable spin-2 tensor: 
\begin{equation}
T_{\mu \nu }=\frac{8}{3}F_{\mu \alpha \beta }^{+}F_{\nu \alpha \beta }^{-}=%
\frac{4}{3}F_{\mu \alpha \beta }F_{\nu \alpha \beta }-\frac{1}{3}\delta
_{\mu \nu }F_{\alpha \beta \gamma }^{2}, 
\label{tt}
\end{equation}
which appears to have the desired properties.
We are going to show that this tensor gives sensible definitions of 
$c$ and $a$ and compute their values.

The unusual factor will be fixed in the next section by 
checking the Poincar\'{e} algebra in the operator-product expansion. 

It is straightforward to show that $T_{\mu \nu }$ is traceless,
gauge-invariant and conserved when the field equations (\ref
{feq}) are satisfied, 
and transforms correctly under coordinate inversion. For the proof
of conservation we observe that the cyclic identity also implies
\[
T_{\mu \nu }=\frac{8}{3}F_{\mu \alpha \beta }^{+}F_{\nu \beta \alpha }^{-}. 
\]
The difference $\Delta $ between the two forms for $T_{\mu \nu }$ is
proportional to $F_{\mu \alpha \beta }^{+}F_{\nu [\beta \alpha ]}^{-}$, the
brackets denoting antisymmetrization. The cyclic identity in (\ref{sym}) can
be expressed as $F_{\mu \nu \alpha }-F_{\mu \alpha \nu }=F_{\alpha \nu \mu }$%
. Similar expressions hold for $\tilde{F}$ and $F^{\pm }$. We have therefore 
$\Delta \sim F_{\mu \alpha \beta }^{+}F_{\alpha \beta \nu }^{-}$. Using the
cyclic identity once more on $F^{+}$ we arrive at $\Delta \propto F_{\alpha
\beta \mu }^{+}F_{\alpha \beta \nu }^{-}=0$.

We are going, with some abuse of language, to call ``stress tensor''
the spin-2 current (\ref{tt}), because this helps us use
formulas from the literature. It is clear, on the other hand,
that it is a fairly new object
and demands a special study. At the moment,
however, I cannot characterize it any better,
and the properties outlined
here are meant to draw attention to it.

\subsection{Computation of $c$}

The field-strength propagator $\langle F_{\mu \nu \alpha }(x)~F_{\rho \sigma
\beta }(0)\rangle $ is, by conformal invariance, $1/|x|^{2d}$ times a linear
combination of the following three conformal structures: 
\begin{eqnarray*}
C_{\mu \nu \alpha ,\rho \sigma \beta }^{(1)}(x) &=&(I_{\mu \rho }(x)I_{\nu
\sigma }(x)-I_{\mu \sigma }(x)I_{\nu \rho }(x))I_{\alpha \beta }(x), \\
C_{\mu \nu \alpha ,\rho \sigma \beta }^{(2)}&=&(I_{\mu \beta }I_{\nu \rho
}-I_{\nu \beta }I_{\mu \rho })I_{\sigma \alpha }-(\rho \leftrightarrow
\sigma ), \\
C_{\mu \nu \alpha ,\rho \sigma \beta }^{(3)} &=&(\delta _{\mu \alpha }I_{\nu
\rho }-\delta _{\nu \alpha }I_{\mu \rho })\delta _{\sigma \beta }-(\rho
\leftrightarrow \sigma ),
\end{eqnarray*}
where $d$ is the dimension of $F$ (2 in the free-field limit). The trace and
cyclic conditions (\ref{sym}) fix the combination of $C^{(1)},$ $C^{(2)}$
and $C^{(3)}$ uniquely up to the overall factor, which can be found by
direct inspection, using the $\chi $-propagator worked out in the previous
section. The final result reads 
\begin{equation}
\langle F_{\mu \nu \alpha }(x)~F_{\rho \sigma \beta }(0)\rangle =\frac{1}{%
2\pi ^{2}}\frac{1}{|x|^{4}}\left( 2~C_{\mu \nu \alpha ,\rho \sigma \beta
}^{(1)}-C_{\mu \nu \alpha ,\rho \sigma \beta }^{(2)}+C_{\mu \nu \alpha ,\rho
\sigma \beta }^{(3)}\right) .  \label{FF}
\end{equation}
A good check is that this correlator satisfies the field equations (\ref{feq}%
).

With (\ref{FF}) we find the
two-point function: 
\[
\langle T_{\mu \nu }(x)~T_{\rho \sigma }(0)\rangle =\frac{4}{45\pi ^{4}}{%
\prod }_{\mu \nu ,\rho \sigma }^{(2)}\left( \frac{1}{|x|^{4}}\right),
\]
and this defines the central charge $c$. We have 
\[
c=\frac{32}{45}.
\]

\subsection{OPE\ structure and computation of $a$}

The OPE\ structure exhibits novel features with respect to the ordinary
theories. In particular, the presence of ghosts is exhibited by higher-spin
composite operators of low dimensionality. The OPE of two stress tensors
contains: the central charge $c$, with singularity $1/|x|^{8}$; the
stress tensor itself, with singularity $1/|x|^{4};$ higher-spin currents of
dimension $2+s,1+s,\cdots 3,$ where $s$ is the spin; descendants and regular
terms. The first higher-spin current is a spin-4, dimension-4 operator
appearing at the same level as the stress tensor (singularity $1/|x|^{4}$).
This operator reads 
\[
{\cal O}_{\mu \nu \rho \sigma }^{(4)}=\sum_{{\rm perms}(\mu \nu \rho \sigma
)}^{\prime }F_{\alpha \mu \nu }^{+}F_{\alpha \rho \sigma }^{-}-{\rm traces}. 
\]
The primed sum is understood to be divided by the number of permutations.
The operator ${\cal O}_{\mu \nu \rho \sigma }^{(4)}$ satisfies the
multiple-conservation condition $\partial _{\mu }\partial _{\nu }\partial
_{\rho }\partial _{\sigma }{\cal O}_{\mu \nu \rho \sigma }^{(4)}=0.$ The
proof of this fact is lengthy and involves repeated use of the cyclic
identity and the field equations. Observe in particular that $\partial _{\mu
}\partial _{\nu }F_{\alpha \mu \nu }=0$ on the solutions to the field
equations. I illustrate the strategy of the proof on the most involved term,
which is 
\[
\partial _{\rho }\partial _{\sigma }F_{\alpha \mu \nu }^{+}\,\partial _{\mu
}\partial _{\nu }F_{\alpha \rho \sigma }^{-}. 
\]
First, we exchange $\mu $ and $\rho $ by using the property of self-duality
in $\alpha \mu $ and anti-self-duality in $\alpha \rho $. We then use the
cyclic identity on $F_{\alpha \mu \sigma }^{-}$ and arrive at 
\[
-\partial _{\rho }\partial _{\sigma }F_{\alpha \rho \nu }^{+}\,\partial
_{\mu }\partial _{\nu }(F_{\mu \sigma \alpha }^{-}+F_{\sigma \alpha \mu
}^{-}). 
\]
We use the field equations to replace $\partial _{\mu }F_{\mu \sigma \alpha
}^{-}$ with $\partial _{\mu }F_{\alpha \mu \sigma }^{-}$ and observe that we
obtain a term identical to the one we started from. 
We move it on the left-hand
side and write 
\[
\partial _{\rho }\partial _{\sigma }F_{\alpha \mu \nu }^{+}\,\partial _{\mu
}\partial _{\nu }F_{\alpha \rho \sigma }^{-}={\frac{1}{2}}\partial _{\rho
}\partial _{\sigma }F_{\alpha \rho \nu }^{+}\,\partial _{\mu }\partial _{\nu
}F_{\alpha \sigma \mu }^{-}. 
\]
Now we use the cyclic identity on $F_{\alpha \rho \nu }^{+}$ and get 
\[
-{\frac{1}{2}}\partial _{\rho }\partial _{\sigma }(F_{\rho \nu \alpha
}^{+}+F_{\nu \alpha \rho }^{+})\,\partial _{\mu }\partial _{\nu }F_{\alpha
\sigma \mu }^{-}. 
\]
Using the field equation $\partial _{\rho }F_{\rho \nu \alpha }^{+}=\partial
_{\rho }F_{\alpha \rho \nu }^{+}$ we finally arrive at 
\[
-{\frac{1}{4}}\partial _{\rho }\partial _{\sigma }F_{\nu \alpha \rho
}^{+}\,\partial _{\mu }\partial _{\nu }F_{\alpha \nu \mu }^{-}=-{\frac{1}{4}}%
\partial _{\rho }\partial _{\sigma }F_{\sigma \alpha \rho }^{+}\,\partial
_{\mu }\partial _{\nu }F_{\alpha \sigma \mu }^{-}=0. 
\]

The other non-trace terms in ${\cal O}_{\mu \nu \rho \sigma }^{(4)}$ can be
shown to vanish in a similar way. Finally, the trace terms always contain
the stress tensor and obey the multiple-conservation condition because the
stress tensor is conserved.

In the basis of \cite{high} we find the OPE expansion 
\begin{eqnarray}
T_{\mu \nu }(x)~T_{\rho \sigma }(0) &=&\frac{4}{45\pi ^{4}}~~{\prod }_{\mu
\nu ,\rho \sigma }^{(2)}\left( \frac{1}{|x|^{4}}\right)  \nonumber \\
&&+\frac{1}{4\pi ^{2}}~T_{\alpha \beta }(0)~\left[ {\rm SP}_{\mu \nu ,\rho
\sigma ;\alpha \beta }\left( \frac{1}{|x|^{2}}\right) +\frac{3}{32}{\prod }%
_{\mu \nu ,\rho \sigma }^{(2)}\partial _{\alpha }\partial _{\beta }\left(
|x|^{2}\ln |x|^{2}\mu ^{2}\right) \right.  \nonumber \\
&&~~~~~~~~~~~~~~~~~~~~\left. -\frac{5}{32}{\prod }_{\mu \nu \alpha ,\beta
\rho \sigma }^{(3)}\left( |x|^{2}\ln |x|^{2}\mu ^{2}\right) \right] 
\nonumber \\
&&+{\frac{1}{4\pi^2}}~{\cal O}^{(4)}_{\alpha\beta\gamma\delta}(0)~ \left[ -{%
\frac{1}{45}}{\prod }_{\mu \nu ,\rho \sigma }^{(2)}\partial _{\alpha
}\partial _{\beta }\partial _{\gamma }\partial _{\delta }\left(|x|^4 \ln
|x|^2 \mu^2 \right) \right.  \nonumber \\
&&~~~~~~~ \left.+{\frac{5}{126}}{\prod }_{\mu \nu \alpha ,\rho \sigma \beta
}^{(3)}\partial _{\gamma }\partial _{\delta }\left(|x|^4 \ln |x|^2 \mu^2
\right)- {\frac{1}{216}} {\prod }_{\mu \nu \rho \sigma ,\alpha \beta \gamma
\delta }^{(4)}\left(|x|^4 \ln |x|^2 \mu^2 \right) \right]  \nonumber \\
&&+{\rm \, less\,\, singular \,\, terms}+{\rm descendants}+{\rm regular\
terms,}  \label{OPE}
\end{eqnarray}
the structure ${\rm SP}_{\mu \nu ,\rho \sigma ;\alpha \beta }\left( \frac{1}{%
|x|^{2}}\right) $ being the generator of the Poincar\'{e} algebra. The
overall coefficient of $T_{\mu \nu }$ has been fixed by matching the
coefficient of $T_{\alpha \beta }{\rm SP}_{\mu \nu ,\rho \sigma ;\alpha
\beta }\left( \frac{1}{|x|^{2}}\right) $ in the OPE, which is universal and
has to be equal to $1/4\pi ^{2}$.

In the calculation of the above OPE, it is necessary to extract the spin-2
content from the product $F_{\mu \nu \alpha}^{+}F_{\rho \sigma\beta
}^{-}+F_{\mu \nu \alpha}^{-}F_{\rho \sigma \beta}^{+}$. It can be proved
that the stress-tensor content of this expression is fixed uniquely by the
symmetry properties in the indices, the cyclic identity, the tracelessness
of $F,$ and relations such as $F_{\alpha \beta \nu }^{+}F_{\alpha \beta
\sigma }^{-}=0,$ $F_{\mu \alpha \beta }^{+}F_{\nu \alpha \beta }^{-}=\frac{3%
}{8}T_{\mu \nu }$, with the result 
\begin{eqnarray}
F_{\mu \nu \alpha}^{+}F_{\rho \sigma \beta}^{-}+F_{\mu \nu\alpha
}^{-}F_{\rho \sigma\beta }^{+}\rightarrow ~~~ {\frac{3}{128}} \left( -2
\delta_{\mu \sigma} \delta_{\nu \rho} T_{\alpha \beta} + 2 \delta_{\mu \rho}
\delta_{\nu \sigma} T_{\alpha \beta} + 3 \delta_{\beta \sigma} \delta_{\nu
\rho} T_{\alpha \mu} - 3 \delta_{\beta \rho} \delta_{\nu \sigma} T_{\alpha
\mu}\right.  \nonumber \\
- 3 \delta_{\beta \sigma} \delta_{\mu \rho} T_{\alpha \nu} + 3 \delta_{\beta
\rho} \delta_{\mu \sigma} T_{\alpha \nu} - \delta_{\beta \nu} \delta_{\mu
\sigma} T_{\alpha \rho} + \delta_{\beta \mu} \delta_{\nu \sigma} T_{\alpha
\rho} + \delta_{\beta \nu} \delta_{\mu \rho} T_{\alpha \sigma} -
\delta_{\beta \mu} \delta_{\nu \rho} T_{\alpha \sigma}- \delta_{\alpha
\sigma} \delta_{\nu \rho} T_{\beta \mu}  \nonumber \\
+ \delta_{\alpha \rho} \delta_{\nu \sigma} T_{\beta \mu} + \delta_{\alpha
\sigma} \delta_{\mu \rho} T_{\beta \nu} - \delta_{\alpha \rho} \delta_{\mu
\sigma} T_{\beta \nu} + 3 \delta_{\alpha \nu} \delta_{\mu \sigma} T_{\beta
\rho} - 3 \delta_{\alpha \mu} \delta_{\nu \sigma} T_{\beta \rho} - 3
\delta_{\alpha \nu} \delta_{\mu \rho} T_{\beta \sigma} + 3 \delta_{\alpha
\mu} \delta_{\nu \rho} T_{\beta \sigma}  \nonumber \\
+ 4 \delta_{\alpha \sigma} \delta_{\beta \nu} T_{\mu \rho} - 4
\delta_{\alpha \nu} \delta_{\beta \sigma} T_{\mu \rho} + 5 \delta_{\alpha
\beta} \delta_{\nu \sigma} T_{\mu \rho}- 4 \delta_{\alpha \rho}
\delta_{\beta \nu} T_{\mu \sigma}+ 4 \delta_{\alpha \nu} \delta_{\beta \rho}
T_{\mu \sigma} - 5 \delta_{\alpha \beta} \delta_{\nu \rho} T_{\mu \sigma} 
\nonumber \\
- 4 \delta_{\alpha \sigma} \delta_{\beta \mu} T_{\nu \rho}+ 4 \delta_{\alpha
\mu} \delta_{\beta \sigma} T_{\nu \rho} - 5 \delta_{\alpha \beta}
\delta_{\mu \sigma} T_{\nu \rho} + 4 \delta_{\alpha \rho} \delta_{\beta \mu}
T_{\nu \sigma} - 4 \delta_{\alpha \mu} \delta_{\beta \rho} T_{\nu \sigma} +
5 \delta_{\alpha \beta} \delta_{\mu \rho} T_{\nu \sigma}\left.\right) . 
\nonumber
\end{eqnarray}
The expression on the left-hand side contains also ${\cal O}%
^{(4)}_{\mu\nu\rho\sigma}$, which is however orthogonal to the stress tensor
and so does not contribute to $c$ and $a$.

We can define our $a$ in the following way. The scalar, spinor and vector
OPE terms $(TT)^T$ are a basis for the OPE structure \cite{high}. We use the
stress-tensor two-point function and the $TT$ OPE to associate effective
numbers $n_{s,f,v} $ of scalars, fermions and vectors to the spin-2
conformal field and then apply the free-fields formulas for $c$ and $a$.

We write 
\[
\langle (TT)\,T\rangle =n_s\langle (TT)\,T\rangle_s+n_f\langle
(TT)\,T\rangle_f +n_v\langle (TT)\,T\rangle_v. 
\]
Here $(TT)$ means that we take the limit in which the distance between the
first two $T$-insertions tends to zero, and so we can use the OPE calculated
above. On the right-hand side, $\langle (TT)\,T\rangle_{s,f,v}$ denote the
corresponding expressions for one free real scalar, one fermion and one
vector, which can be read in \cite{high}. Clearly, only the $T$-content of
the OPE is relevant in the limit we are considering: $\langle (TT)\,
T\rangle =(TT)^T\langle T\,T \rangle$, where $(TT)^T$ denotes the structure
multiplying $T$ in the $TT$ OPE. For example, $(TT)^T$ is the structure
contained between the first square brackets in (\ref{OPE}). We have 
\begin{equation}
c\, (TT)^T={\frac{1}{120}}\left[n_s (TT)^T_s+6\, n_f (TT)^T_f +12 \, n_v
(TT)^T_v \right].  \label{YU}
\end{equation}

Using the two-point functions and OPEs of free fields \cite{high} we arrive,
by comparison, at 
\[
n_{s}=0,\qquad \qquad n_{f}={\frac{256}{27}},\qquad \qquad n_{v}={\frac{64}{%
27}}. 
\]
Observe that $n_{s}=0$ can be inferred immediately from the OPE. Scalar
fields produce a structure $(TT)_{s}^{T}$ with the maximal number of
uncontracted $x_{\mu }$'s (six), vector fields give a structure $%
(TT)_{v}^{T} $ with the minimum number (two) and $(TT)_{f}^{T}$, for the
spinors, contain four uncontracted $x_{\mu }$'s. A quick inspection of the
propagator shows that our structure $(TT)^{T}$ cannot contain more than four
uncontracted $x_{\mu }$'s.

The final result is 
\[
c={\frac{32}{45}},\qquad \qquad a={\frac{848}{1215}},\qquad \qquad {\frac{c-a%
}{c}}={\frac{1}{54}}. 
\]
We see that both $c$ and $a$ are positive, as well as $n_{f}$ and $n_{v}$,
and that $c$ is ``almost'' equal to $a$, but slightly greater.

The procedure used to calculate $c$ and $a$ (\ref{YU}) guarantees that these
values parametrize the trace anomaly in the appropriate way. However, we
cannot write for the trace anomaly a closed expression such as (\ref{1.1}),
which makes use of the coupling to external gravity, and we need to work
always at the level of correlators and OPEs. It is meaningful, nevertheless,
to truncate the right-hand side of (\ref{1.1}) to the quadratic terms in an
expansion of the gravitational field around flat space.

We have therefore shown that $c$ and $a$ can be appropriately defined in our
theories despite the absence of a coupling to external gravity, and that
they are positive. Some issues need to be better understood, for example the
relation (if any) between the gauge-invariant tensor $T_{\mu \nu }$ and
the Noether tensor.

The ${\cal O}_{\alpha \beta \gamma \delta }^{(4)}$-content of the OPE can be
extracted with the replacement 
\[
F_{\mu \nu \alpha }^{+}F_{\rho \sigma \beta }^{-}+F_{\mu \nu \alpha
}^{-}F_{\rho \sigma \beta }^{+}\rightarrow \delta _{\nu \sigma }{\cal O}%
_{\alpha \beta \mu \rho }^{(4)}-\delta _{\nu \rho }{\cal O}_{\alpha \beta
\mu \sigma }^{(4)}-\delta _{\mu \sigma }{\cal O}_{\alpha \beta \nu \rho
}^{(4)}+\delta _{\mu \rho }{\cal O}_{\alpha \beta \nu \sigma }^{(4)}. 
\]
The presence of this multiply-conserved, spin-4, dimension-4 operator,
absent in ordinary theories, is here emphasized, as a good illustration of
the new features of higher-spin conformal field theory and the role of the
multiple-conservation condition. The hope is that the ghost degrees of
freedom, or spin-$s$ operators with dimension lower than $2+s$, might be
controlled in some way. A sufficiently strong interaction might raise the
dimensions of all operators. I recall that the Nachtmann theorem \cite{nach}%
, in unitary theories, states that the anomalous dimensions of the
higher-spin currents generated by the singular terms of the OPE are to some
extent correlated \cite{n=4,n=2} (e.g. the anomalous dimensions increase
with the spin and the magnitude of the interaction). It is conceivable that
a similar result here would ensure that below a certain energy threshold,
when the interaction is sufficiently strong, the theory is perfectly
unitary, i.e. all spin-$s$ operators have dimension greater than or equal to 
$2+s$.

\subsection{Antisymmetric conformal tensors}

With antisymmetric tensors, many of the nice features of symmetric tensors
disappear. In particular, conformal invariance spoils both the positivity of
the action and gauge invariance. A 2-form $A_{\mu \nu }$ has the
conformal-invariant action 
\[
S=\frac{1}{2}\int \left[ (\partial _{\alpha }A_{\mu \nu })^{2}-4(\partial
_{\alpha }A_{\alpha \nu })^{2}\right] . 
\]
With $A_{\mu \nu }=\partial _{\mu }\zeta _{\nu }-\partial _{\nu }\zeta _{\mu
}$ we find $S=-\frac{1}{2}\int (\partial _{\alpha }A_{\mu \nu })^{2}$, so
that the action is not positive-definite and gauge invariance is completely
lost. The theory can be coupled in a (classically) conformal way to Abelian
and non-Abelian gauge fields, as well as gravity. Renormalizable couplings
to symmetric higher-spin conformal fields is instead problematic. For
example, a coupling of a complex antisymmetric tensor with a spin-3 field of
the Pauli type, such as 
\[
igF_{\mu \nu \alpha \beta }A_{\mu \alpha }\bar{A}_{\nu \beta }={\cal O}_{\mu
\nu \rho }^{(3)}\chi _{\mu \nu \rho }+{\rm total\,\, derivatives} 
\]
vanishes because of the cyclic identity.

The $A_{\mu\nu}$-field equations and propagator read 
\[
\Box A_{\mu\nu}-2\partial_\alpha(\partial_\mu A_{\alpha\nu}+\partial_\nu
A_{\mu\alpha} )=0,\qquad \langle A_{\mu\nu}(x)\, A_{\rho\sigma}(0)\rangle={%
\frac{-1}{8\pi^2 |x|^2}} \left(
I_{\mu\rho}(x)I_{\nu\sigma}(x)-I_{\mu\sigma}(x)I_{\nu\rho}(x) \right). 
\]
Observe that the propagator is reflection-negative. We conclude that
antisymmetric conformal tensor fields are much less interesting than the
symmetric tensors.

\section{Conformal fermionic fields}

A spin-$\left( s+1/2\right) $ field is described by a spinor $\psi _{\mu
_{1}\cdots \mu _{s}}$ with $s$ Lorentz indices, completely symmetric and
traceless.

The transformation of the spinor under coordinate inversion is 
\[
\psi _{\mu _{1}\cdots \mu _{s}}\rightarrow |x|^{2}x\!\!\!\slash\gamma
_{5}I_{\mu _{1}}^{\nu _{1}}(x)\cdots I_{\mu _{s}}^{\nu _{s}}(x)\psi _{\nu
_{1}\cdots \nu _{s}}. 
\]
The contraction $\gamma _{\beta }\psi _{\beta \mu _{2}\cdots \mu _{s}}$
transforms as a spin-$\left( s-1/2\right) $ conformal spinor. Further
contractions with gamma matrices are automatically zero, owing to complete
tracelessness. Instead $\sum_{i=1}^{s}\gamma _{\mu _{i}}\gamma _{\alpha
}\psi _{\alpha \mu _{1}\cdots \hat{\mu}_{i}\cdots \mu _{s}}$ transforms as a
spin-$\left( s+1/2\right) $ spinor. Therefore we can always impose 
\begin{equation}
\gamma _{\alpha }\psi _{\alpha \mu _{2}\cdots \mu _{s}}=0  \label{tra}
\end{equation}
and preserve conformal invariance. Under this condition the most general
conformal lagrangian is simply 
\begin{equation}
{\cal L=}\overline{\psi }_{\mu _{1}\cdots \mu _{s}}\partial \!\!\!\slash %
\psi _{\mu _{1}\cdots \mu _{s}},  \label{hoo}
\end{equation}
any other possible term vanishing because of (\ref{tra}). The proof that (%
\ref{hoo}) transforms correctly is rather lengthy, but straightforward. To
make (\ref{tra}) manifest, we can insert appropriate projectors: 
\begin{eqnarray*}
{\cal L} &=&\left( \overline{\psi }_{\mu _{1}\cdots \mu _{s}}-\frac{1}{2(s+1)%
}\sum_{i=1}^{s}\overline{\psi }_{\alpha \mu _{1}\cdots \hat{\mu}_{i}\cdots
\mu _{s}}\gamma _{\alpha }\gamma _{\mu _{i}}\right) \partial \!\!\!\slash %
\left( \psi _{\mu _{1}\cdots \mu _{s}}-\frac{1}{2(s+1)}\sum_{i=1}^{s}\gamma
_{\mu _{i}}\gamma _{\alpha }\psi _{\alpha \mu _{1}\cdots \hat{\mu}_{i}\cdots
\mu _{s}}\right) = \\
&=&\overline{\psi }_{\mu _{1}\cdots \mu _{s}}\partial \!\!\!\slash \psi
_{\mu _{1}\cdots \mu _{s}}-\frac{s}{s+1}\overline{\psi }_{\alpha \mu
_{2}\cdots \mu _{s}}\gamma _{\alpha }\partial _{\beta }\psi _{\beta \mu
_{2}\cdots \mu _{s}}-\frac{s}{s+1}\overline{\psi }_{\alpha \mu _{2}\cdots
\mu _{s}}\gamma _{\beta }\partial _{\alpha }\psi _{\beta \mu _{2}\cdots \mu
_{s}} \\
&&+\frac{s(s+2)}{2(s+1)^{2}}\overline{\psi }_{\alpha \mu _{2}\cdots \mu
_{s}}\gamma _{\alpha }\partial \!\!\!\slash \gamma _{\beta }\psi _{\beta \mu
_{2}\cdots \mu _{s}}.
\end{eqnarray*}

The field equations are 
\[
\partial \!\!\!\slash \psi _{\mu _{1}\cdots \mu _{s}}=\frac{1}{s+1}%
\sum_{i=1}^{s}\gamma _{\mu _{i}}\partial _{\alpha }\psi _{\alpha \mu
_{1}\cdots \hat{\mu}_{i}\cdots \mu _{s}}. 
\]

Condition (\ref{tra}) is not sufficient to eliminate the ghosts of the
theory. We see that no gauge invariance survives and the theory can be
straightforwardly coupled to Abelian and non-Abelian gauge fields, as well
as gravity. In particular, $c$ and $a$ can be defined in the usual way. In
the next section, I discuss the case $s=1$ in detail and compute the
contribution of conformal spinors to the gauge beta function.

\subsection{Spin 3/2}

For $s=1$ the action 
\begin{eqnarray*}
S &=&\int {\cal L}=\int \overline{\psi }_{\mu }\left[ \partial \!\!\!\slash %
\psi _{\mu }-\frac{1}{2}\gamma _{\alpha }\partial _{\mu }\psi _{\alpha }-%
\frac{1}{2}\gamma _{\mu }\partial _{\alpha }\psi _{\alpha }+\frac{3}{8}%
\gamma _{\mu }\partial \!\!\!\slash \gamma _{\alpha }\psi _{\alpha }\right] =
\\
&=&\int \overline{\psi }_{\mu }P_{\mu \nu }\partial \!\!\!\slash P_{\nu \rho
}\psi _{\rho },\qquad \qquad P_{\mu \nu }=\delta _{\mu \nu }-\frac{1}{4}%
\gamma _{\mu }\gamma _{\nu }
\end{eqnarray*}
is invariant under coordinate inversion, the field being transformed as 
\[
\psi _{\mu }\rightarrow |x|^{2}x\!\!\!\slash\gamma _{5}I_{\mu }^{\nu
}(x)\psi _{\nu }. 
\]
The field equations are 
\begin{equation}
\partial \!\!\!\slash \psi _{\mu }=\frac{1}{2}\gamma _{\mu }\partial \cdot
\psi ,  \label{eqa}
\end{equation}
bearing in mind that $\gamma \cdot \psi =0$. The field equations imply also 
\[
\left( \Box \delta _{\mu \nu }-\partial _{\mu }\partial _{\nu }\right) \psi
_{\nu }=0,\qquad \qquad \partial \!\!\!\slash \partial \cdot \psi =0. 
\]
The transversal component of $\psi _{\mu }$ obeys an ordinary wave equation,
while $\partial \cdot \psi $ obeys the Dirac equation. The transformation $%
\delta \psi _{\mu }=\partial _{\mu }\epsilon $ is not a symmetry, however,
since it preserves neither $\gamma \cdot \psi =0$ nor (\ref{eqa}).

Our theory coincides with the theory called ``singular'' by Haberzett in the
context of the nuclear theory of hadronic resonances: see formula (40) of
ref. \cite{haberzett}. Its conformal invariance, and the unicity of the
theory in this respect, is here emphasized.

I\ investigate in detail the coupling to Abelian and non-Abelian gauge
fields, obtained by covariantizing the derivatives: 
\[
{\cal L}=\frac{1}{4}(F_{\mu \nu }^{a})^{2}+\overline{\psi }_{\mu }^{i}\left[
D\!\!\!\!\slash^{ij}\psi _{\mu }^{j}-\frac{1}{2}\gamma _{\alpha }D_{\mu
}^{ij}\psi _{\alpha }^{j}-\frac{1}{2}\gamma _{\mu }D_{\alpha }^{ij}\psi
_{\alpha }^{j}+\frac{3}{8}\gamma _{\mu }D\!\!\!\!\slash^{ij}\gamma _{\alpha
}\psi _{\alpha }\right] . 
\]
Here $a$ is the index of the fundamental representation of the gauge group $%
G $, and $i,j$ are indices of the matter representation $R$. The notation for
the covariant derivative is $D_{\mu }^{ij}\psi _{\nu }^{j}=\partial _{\mu
}\psi _{\nu }^{i}+g(T^{a})^{ij}A_{\mu }^{a}\psi _{\nu }^{j},$ as usual. The
spin-3/2 propagator is 
\begin{eqnarray*}
\langle \psi _{\mu }^{i}(k)~\overline{\psi }_{\nu }^{j}(-k)\rangle &=&-\frac{%
i\delta ^{ij}}{k^{2}}\left[ k\!\!\!\slash\delta _{\mu \nu }-k_{\mu }\gamma
_{\nu }-k_{\nu }\gamma _{\mu }+\frac{1}{2}\gamma _{\mu }k\!\!\!\slash\gamma
_{\nu }+\frac{2}{k^{2}}k_{\mu }k\!\!\!\slash k_{\nu }\right] \\
&=&-\frac{i\delta ^{ij}}{k^{2}}P_{\mu \alpha }k\!\!\!\slash\left( \delta
_{\alpha \beta }+2\frac{k_{\alpha }k_{\beta }}{k^{2}}\right) P_{\beta \nu }
\end{eqnarray*}
and the vertex is 
\[
\langle \psi _{\mu }^{i}~\overline{\psi }_{\nu }^{j}~A_{\rho }^{a}\rangle
=-gT_{ij}^{a}P_{\mu \alpha }\gamma _{\rho }P_{\alpha \nu }. 
\]

The theory is conformal at the classical level, and scale invariance is
broken, as usual, by the radiative corrections at the quantum level. I have
computed the one-loop beta function of this model in two different ways
(gluon self-energy and three-gluon vertex), with the result 
\[
\beta (g)=-\frac{g^{3}}{48\pi ^{2}}\left[
11C(G)-20C(R_{3/2})-4C(R_{1/2})\right] . 
\]
The correction due to our spin-3/2 field is the term proportional to $%
C(R_{3/2})$, while the term proportional to $C(R_{1/2})$ is the usual
spin-1/2 contribution, here inserted for comparison.

We see that this peculiar type of ``matter'' contributes to the beta
function with the same sign as ordinary matter. For $C(R_{3/2})\lesssim 
\frac{11}{20}C(G)$ the one-loop beta function is arbitrarily small with
respect to the higher-order corrections, which allow us to conclude that
there is a non-trivial IR fixed point, trustable in perturbation theory, and
a conformal window, which is the main reason why these theories are an
interesting laboratory of models for the ideas of \cite{proc}. Similar
arguments extend to arbitrary half-integer spin. Note that our theories, in
spite of their non-unitarity, are renormalizable and are not
higher-derivative. For this reason we do not compare our results, for
example, with the supergravity calculations,
which cannot give evidence of a conformal window.

\section{Conclusions}

We have explored 
higher-derivative theories and higher-spin conformal theories, 
and studied the central
charges $c$ and $a$, their positivity properties, the existence
of renormalizable couplings and conformal windows.
These theories are good
toy models for investigations in the spirit of \cite{proc,cea}, 
the study of questions concerning quantum irreversibility
and the search for a unified description 
of higher- and lower-spin fields.
Antisymmetric conformal tensors and
higher-derivative theories exhibit severe violations of 
positive-definiteness and reflection positivity. 
Instead, there is evidence that
higher-spin conformal symmetric tensors and fermions
have positive $c$ and $a$. The symmetric tensors, moreover,
have a positive-definite action
and a peculiar gauge symmetry.
These properties, I believe, make higher-spin conformal theories
worthy of attention, even if they are not unitary.

\medskip 
{\bf Acknowledgement}. I am grateful to U. Aglietti, F. Bastianelli, H.B.
Nielsen, R. Rattazzi and A. Waldron for discussions.

\end{document}